\documentclass[twocolumn]{aastex62}

\usepackage{apjfonts}

\shorttitle{Nebula around ASASSN-19\MakeLowercase{ds}}
\shortauthors{Bond et al.}

\newcommand{\CIII}{\ion{C}{3}}

\newcommand{\Ha}{H$\alpha$}

\newcommand{\NIII}{\ion{N}{3}}
\newcommand{\OIII}{\ion{O}{3}}

\newcommand{\kms}{{\>\rm km\>s^{-1}}}
\newcommand{\masyr}{\rm mas\, yr^{-1}}

\def\hei{\ion{He}{1}}
\def\heii{\ion{He}{2}}

\def\nii{\ion{N}{2}}

\def\oiii{\ion{O}{3}}

\def\Gaia{{\it Gaia}}
\newcommand{\GALEX}{{\it GALEX}}

\newcommand{\TESS}{{\it TESS}}

\def\AS{ASASSN-19ds}

\begin{document}

\title{Discovery of Faint Nebulosity Around a Z~Camelopardalis-type Cataclysmic Variable in Antlia: Nova Shell or Ancient~Planetary~Nebula?}

\author[0000-0003-1377-7145]{Howard E. Bond}
\affil{Department of Astronomy \& Astrophysics, Penn State University, University Park, PA 16802, USA}
\affil{Space Telescope Science Institute, 
3700 San Martin Dr.,
Baltimore, MD 21218, USA}





\author[0000-0002-9018-9606]{Dana Patchick}
\affil{Deep Sky Hunters Consortium, 1942 Butler Ave. Los Angeles, CA 90025, USA }

\author[0009-0005-9964-1602]{Daniel Stern}
\affil{MEA Observatory,
16252 Andalucia Ln., 
Delray Beach, FL  33446, USA }

\author[0009-0009-3986-4336]{Jonathan Talbot}
\affil{Stark Bayou Observatory, 1013 Conely Cir., Ocean Springs, MS 39564, USA}

\author[0000-0002-4964-4144]{John R. Thorstensen}
\affil{Department of Physics \& Astronomy, 6127 Wilder Laboratory, Dartmouth College, Hanover, NH 03755, USA}


\correspondingauthor{Howard E. Bond}
\email{heb11@psu.edu}

\begin{abstract}

We report our discovery of a faint nebula surrounding a previously little-studied 15th-mag variable star, \AS, in the southern-hemisphere constellation Antlia. Spectra verify that the star is a  cataclysmic variable (CV). Using new and archival photometry, we confirm that \AS\ is an eclipsing binary with an orbital  period of 0.139~days (3.34~hr). Moreover, its out-of-eclipse brightness shows a ``sawtooth'' light curve with an amplitude of $\sim$1~mag and an interval between peaks that varies between about three to nearly five weeks. Its mean absolute magnitude in the \Gaia\/ system is $M_G=+6.5$. These combined properties lead to a classification of \AS\ as a Z~Camelopardalis-type CV\null. We obtained deep narrow-band images of the nebulosity, using modest-sized telescopes and extremely long exposure times. Our imagery reveals a bipolar morphology, with thin arcs at the ends of the major axis, likely indicating an interaction with the interstellar medium. We consider several scenarios for the origin of the nebula, but from the existing information we are unable to distinguish between it being ejecta from an unobserved classical-nova outburst several centuries ago, or an ancient planetary nebula. Future observations should be able to decide between these possibilities. At the star's distance of only $\sim$557~pc, a nova eruption would have been a spectacular naked-eye event.

\null\vskip 0.2in

\end{abstract}



\section{Introduction}

\subsection{Faint Nebulae Around Cataclysmic Variable Stars \label{subsec:faint_nebulae} }

Cataclysmic variables (CVs) are close binaries in which a Roche-lobe--filling star transfers mass to a compact companion, which in most cases is a white dwarf (WD)\null. Unless the WD has a very strong magnetic field, the transferred gas forms an accretion disk, from which most of the material eventually falls onto the WD, although some of it leaves the system as a fast wind. The major classes of CVs are classical novae (CNe, in which the accreted hydrogen accumulates until it ignites nuclear fusion on the surface of the WD); dwarf novae (DNe, in which the accretion disk is usually optically thin, but becomes optically thick and brighter during occasional outbursts); and nova-like variables (NLVs, in which the transfer rate is so high that the accretion disk remains optically thick most or all of the time). The Z~Camelopardalis variables \citep[ZCVs; see the review by][]{Simonson2014} are a subset of the DNe. They exhibit eruptions similar to those of DNe, but can occasionally remain in ``standstills'' at an intermediate brightness level for a few days up to many years. For reviews of CVs, see the monographs and papers by \citet{Warner1995}, \citet{Szkody2012}, and \citet{Sion2023}.


Following the eruption of a CN, material ejected from the explosion forms an expanding nebula, which typically remains visible for decades to a century or more, before dissipating into the interstellar medium (ISM). A few cases of very faint nebulae centered on NLVs and ZCVs have been discovered, often showing a hollow and filamentary structure. Two examples are the nebulosities around Z~Cam itself \citep{Shara2007, SharaZCam2012, Shara2024} and the ZCV AT~Cancri \citep{SharaATCnc2012}. These nebulae are plausibly attributed to ejection from CN outbursts of the central binaries that occurred several centuries to more than a millennium ago. Aside from these nova shells, nebulae around CVs are remarkably rare, in spite of several extensive searches which have revealed only about another half-dozen examples \citep[e.g.,][and references therein]{Tappert2020} . We reviewed the literature on this subject in a recent paper \citep[][hereafter Paper~I]{BondSYCnc2024}, which gives further details and references.


Advanced amateur astronomers are able to use modest-sized telescopes equipped with modern instrumentation to accumulate extremely long exposure times, for the detection of low-surface-brightness nebulae. In Paper~I we described our serendipitous discovery of a very faint nebula around the ZCV SY~Cancri. This nebula is unusual because the star is off-center, lying near the edge of the nebula, where it is accompanied by a bow shock. We argued that, unlike the ejecta from novae and those around Z~Cam and AT~Cnc, the SY~Cnc nebula is the result of a chance high-speed encounter between a NLV and an interstellar cloud. Our picture is that a fast wind launched from the binary's accretion disk is responsible for the bow shock. The system leaves an off-center Str\"omgren zone in its wake, recombining after being photoionized by the CV's ultraviolet and X-ray radiation and collisionally excited by the fast wind.

In a subsequent paper \citep[][hereafter Paper~II]{BondLSPeg2025}, we presented discoveries of two more cases of off-center \Ha\ nebulae and bow shocks, associated with the NLVs LS~Pegasi and ASASSN-V J205457.73+515731.9. These objects join V341~Arae and BZ~Camelopardalis in this class of CVs interacting with the ISM\null. As discussed in Paper~II, all five binaries have relatively bright absolute magnitudes in the narrow range $+4.5<M_G<+4.8$, consistent with the luminous accretion disks required to launch the winds and produce bow shocks.

\subsection{Discovery and Confirmation of a Nebula around Cataclysmic Variable ASASSN-19ds}

As a follow-up to these findings, one of us (Patchick) has been carrying out a systematic search for faint nebulae associated with known CVs, based on publicly available sky-survey images. Targets are chosen from the catalog\footnote{\url{https://vizier.cds.unistra.fr/viz-bin/VizieR-3?-source=J/AJ/165/163/table1}} of 1,587 CVs assembled by \citet{Canbay2023}. In the course of his survey, Patchick noted a previously uncataloged low-surface-brightness nebula, with a major axis of very roughly $3'$, which appeared to surround the little-studied CV ASASSN-19ds. The discovery was made on the red-sensitive photographic image of the object in the Space Telescope Science Institute Digitized Sky Survey.\footnote{\url{https://archive.stsci.edu/cgi-bin/dss_form}} 


The reality of this very faint nebula appeared to be supported through inspection of wide-angle narrow-band \Ha\ images of the site from the Southern H-Alpha Sky Survey Atlas\footnote{\url{https://amundsen.swarthmore.edu/SHASSA}} (SHASSA) project \citep{Gaustad2001}. Patchick's discovery was made in 2024 November, and communicated shortly thereafter to advanced amateur coauthors Talbot and Stern. 

Exploratory confirmation was obtained on 2025 January~28--30 by Talbot, who accumulated 10.2~hours of exposure time on the object (61~exposures of 600~s each) with his portable 80~mm $f/6$ refractor from a site on West Summerland Key, Florida, at latitude $24^\circ39'$. The telescope was equipped with a ZWO ASI2600MC Pro CMOS camera\footnote{\url{https://www.zwoastro.com/product/asi2600}} and an Antlia ALP-T dual-band \Ha+[\OIII] filter.\footnote{\url{http://www.antliafilter.com}} The frames were stacked into a final image, using PixInsight\footnote{\url{https://pixinsight.com}} v.~1.8.9, and the DBXtract script to extract the \Ha\ signal.

\begin{figure}[h]
\centering
\includegraphics[width=0.47\textwidth]{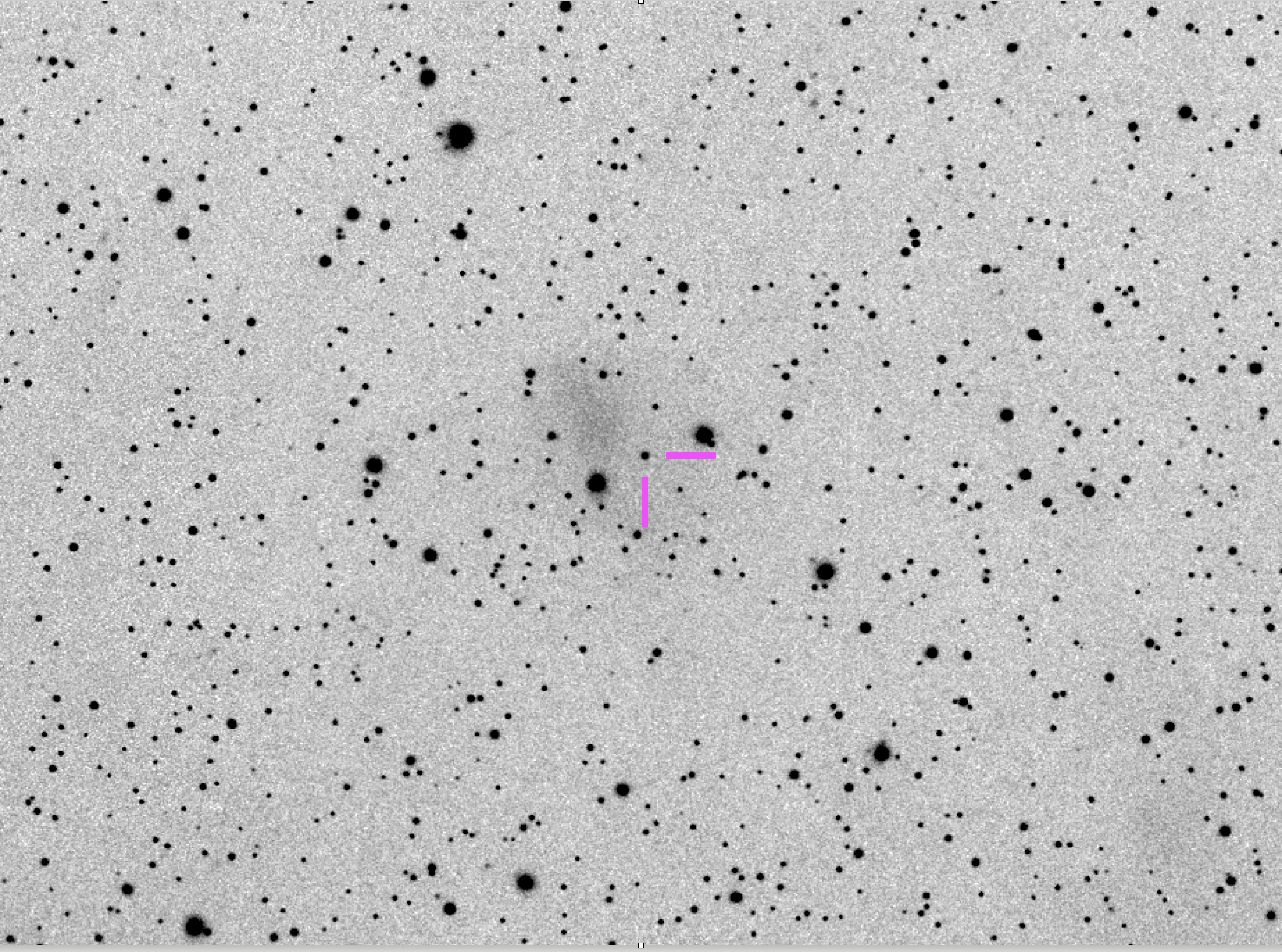}
\caption{
\Ha\ image of the nebula around ASASSN-19ds. Height of frame is $18'$; north is at the top, east on the left. Magenta lines mark the cataclysmic variable. 
\label{fig:Talbot_image}
}
\end{figure}

Figure~\ref{fig:Talbot_image} shows the resulting \Ha\ image, with \AS\ marked at the center. Faint \Ha\ emission surrounds the star, but is brightest on the northeast side. Little or no [\OIII] emission was detected at this exposure level; not surprisingly, since the target was at a mean altitude of only $23^\circ$ above the horizon and the observing site is at sea level. 

In the following section, we present an overview of the basic properties of ASASSN-19ds, which we hope will encourage more detailed investigations of this object. We then present follow-up deep imaging of the nebula obtained by Stern at an excellent southern-hemisphere site. Then we describe several possible scenarios for the origin of the faint nebulosity. We conclude with a summary of our findings, and recommendations for follow-up studies that would be helpful in deciding between the scenarios. 

\null\smallbreak

\section{The Cataclysmic Variable ASASSN-19\MakeLowercase{ds}}

\subsection{Discovery of Variability and Eclipses}

A new 15th-mag variable star in the southern constellation Antlia, designated ASASSN-19ds, was discovered in the course of the All-Sky Automated Survey for SuperNovae \citep[ASAS-SN;][]{Kochanek2017}. It was announced as a candidate CV in 2019~February on the ASAS-SN Transients website.\footnote{\citet{Shappee2014}; \url{https://www.astronomy.ohio-state.edu/asassn/transients.html}}

The discovery was followed up immediately with photometry by two amateurs. B.~Monard, on 2019 March~3, detected\footnote{\url{http://ooruri.kusastro.kyoto-u.ac.jp/mailarchive/vsnet-alert/23040}} an eclipse. Three days later, the eclipses were confirmed\footnote{\url{http://ooruri.kusastro.kyoto-u.ac.jp/mailarchive/vsnet-alert/23041}} by J.~Hambsch based on four nights of data, finding a period of 0.139~days (3.34~hours). The same day T.~Kato reported\footnote{\url{http://ooruri.kusastro.kyoto-u.ac.jp/mailarchive/vsnet-alert/23044}} that he had examined the available ASAS-SN data, allowing him to refine the eclipse period to 
$0.139035164 \pm 0.000000012$~days. Since this flurry of activity, the object has attracted little further attention, up until the exploratory findings and observations presented in this section. It was, however, included in catalogs of stars considered to be hot subdwarfs, based on searches of \Gaia\/ data releases by \citet{Geier2019}, \citet{ Barlow2022}, and \citet{Culpan2022}. 

Table~\ref{tab:DR3data} lists data for ASASSN-19ds, taken from \Gaia\/ Data Release~3\footnote{\url{https://vizier.cds.unistra.fr/viz-bin/VizieR-3?-source=I/355/gaiadr3}} (DR3; \citealt{Gaia2016, Gaia2023}). The bottom row gives its nominal absolute magnitude in the \Gaia\/ system. This was determined using a distance of $557.1^{+7.3}_{-9.2}$~pc from \citet{BailerJones2021}, and corrected for an interstellar reddening of $E(B-V)=0.083$ at the distance of the star, estimated using the online {\tt GALExtin} tool.\footnote{\citet{Amores2021}; \url{http://www.galextin.org/}} 
{ The object was also detected in the ultraviolet by the {\it Galaxy Evolution Explorer\/} (\GALEX), yielding magnitudes of $m_{\rm FUV}=16.36$ and $m_{\rm NUV}=16.20$ \citep{Bianchi2017}. }

\begin{deluxetable}{lc}[b]
\tablecaption{\Gaia\/ DR3 Data for ASASSN-19ds \label{tab:DR3data} }
\tablehead{
\colhead{Parameter}
&\colhead{Value}
}
\decimals
\startdata
RA (J2000)  & 11 05 29.505  \\
Dec (J2000) &  $-40$ 07 08.64  \\
$l$ [deg] &  281.87  \\
$b$ [deg] &  +18.38 \\
Parallax [mas] & $1.7616  \pm 0.0293 $   \\
$\mu_\alpha$ [mas\,yr$^{-1}$] & $-13.662  \pm 0.029 $  \\
$\mu_\delta$ [mas\,yr$^{-1}$] & $ 1.215 \pm 0.031 $   \\
$G$ [mag] & 15.33  \\
$G_{\rm BP}-G_{\rm RP}$ [mag] & 0.35  \\
$M_G$ [mag] & +6.5  \\
\enddata
\end{deluxetable}


\subsection{Spectroscopy}

\begin{figure}[h]
\centering
\includegraphics[width=0.47\textwidth]{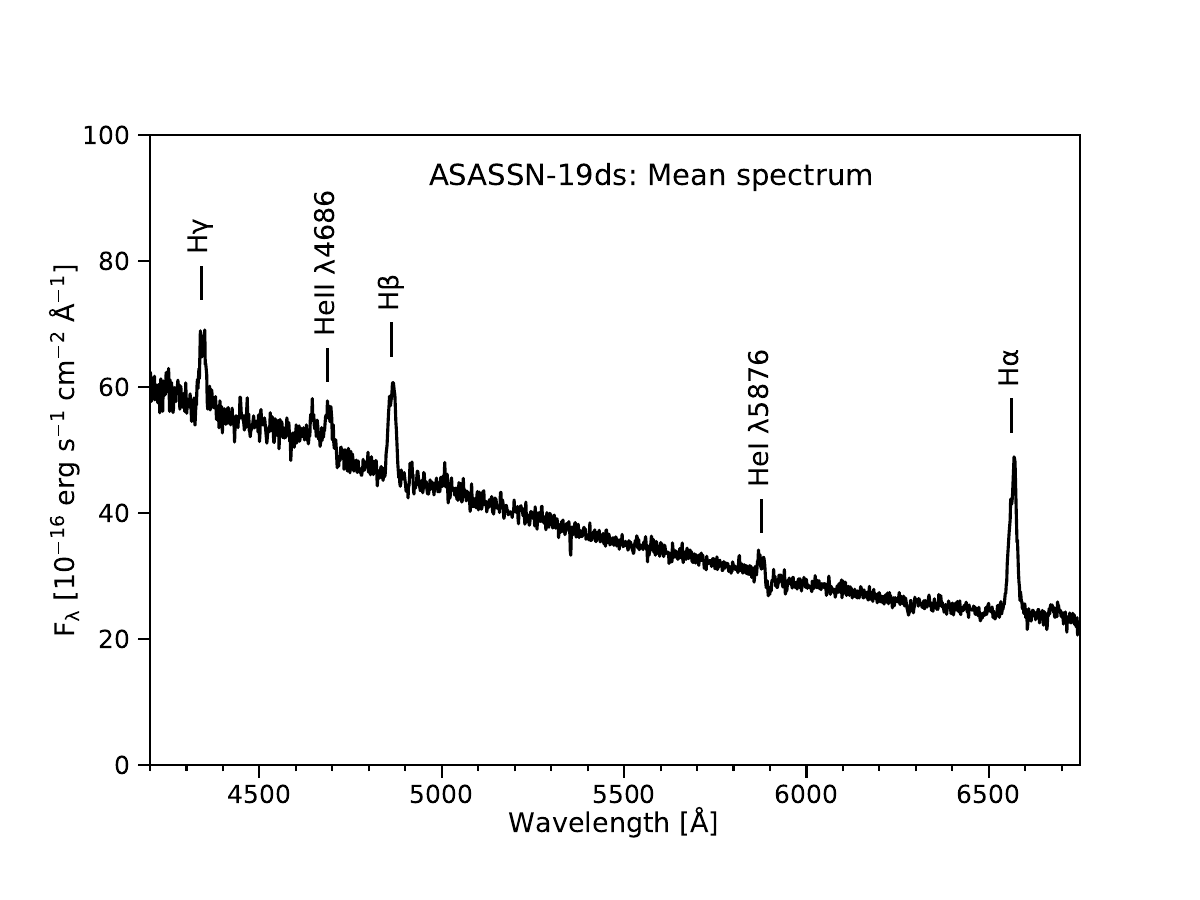}
\vskip-0.2in
\caption{
SAAO spectrum of \AS, showing strong emission lines of hydrogen and \hei\ and \heii, and a feature near $\lambda$4640 due to \CIII\ and \NIII\null. The spectrum confirms that \AS\ is a cataclysmic variable.
\label{fig:saaospectrum}
}
\end{figure}

Coauthor Thorstensen obtained spectroscopy and photometry of ASASSN-19ds during a visit to the South African Astronomical Observatory (SAAO) with a team of undergraduates from Dartmouth College (see acknowledgments). The
spectra are from the SpUpNIC spectrograph \citep{Crause2019},
mounted on the 1.9 m Radcliffe Telescope
at SAAO\null. Grating~6 was used with a $1\farcs35$ slit, covering
4175 to  6780~\AA\  with
4.5~\AA\ spectral resolution.
The spectra were flux-calibrated using observations of
standard stars from \citet{Hamuy1992}.
Figure~\ref{fig:saaospectrum} plots the average of three 720~s exposures,
obtained on 2025 February~17.

The spectrum shows a strong blue continuum with prominent
emission lines. There is strong emission at the Balmer series, with
\Ha\ having an equivalent width of
$\sim$22~\AA, and a FWHM of about 23~\AA\null.  \hei\ $\lambda5876$ and \heii\
$\lambda$4686 also appear in emission, along with a
blend near $\lambda$4640, usually attributed
to \CIII\ and \NIII\null. The observation confirms the classification of the object as a CV\null. { The spectrum shown in Figure~\ref{fig:saaospectrum} has been sky-subtracted, and shows no emission lines from the surrounding nebula. We inspected the sky spectra and found no trace of the nebular spectrum, which is not surprising because the nebula is extremely faint, and the exposures were relatively short.}

\subsection{Time-Series Photometry}

Time-series photometry was obtained on two
nights at SAAO (2025 February~19 and~22), using the Sutherland High-Speed Optical
Camera (SHOC; \citealt{Coppejans2013}) mounted on the 1.0 m telescope.
Individual exposures were 30~s, with
negligible deadtime between exposures, and
all data were taken without a filter.  We subtracted
an average bias from the frames and divided by
flat-field images taken in twilight.
The photometry reduction program used the {\tt IRAF}\footnote{{\tt IRAF} and {\tt PyRAF} were distributed by the National Optical Astronomy Observatories, operated by AURA, Inc., under cooperative agreement with the National Science Foundation.} {\tt digiphot} task, called from {\tt python}
using {\tt PyRAF}\null.  Because the guiding often
drifted, the software centroided the stars on the
individual frames and measured instrumental magnitudes
in a $4''$ diameter software aperture.
Differential magnitudes were taken with respect
to a comparison star lying $69''$ from the target
at position angle $129^\circ$, for which 
\Gaia\/ DR3 lists an apparent magnitude of $G = 11.05$.  We added this to
the differential magnitudes, converting
them to a rough magnitude scale. 

\begin{figure}[h]
\centering
\includegraphics[width=0.47\textwidth]{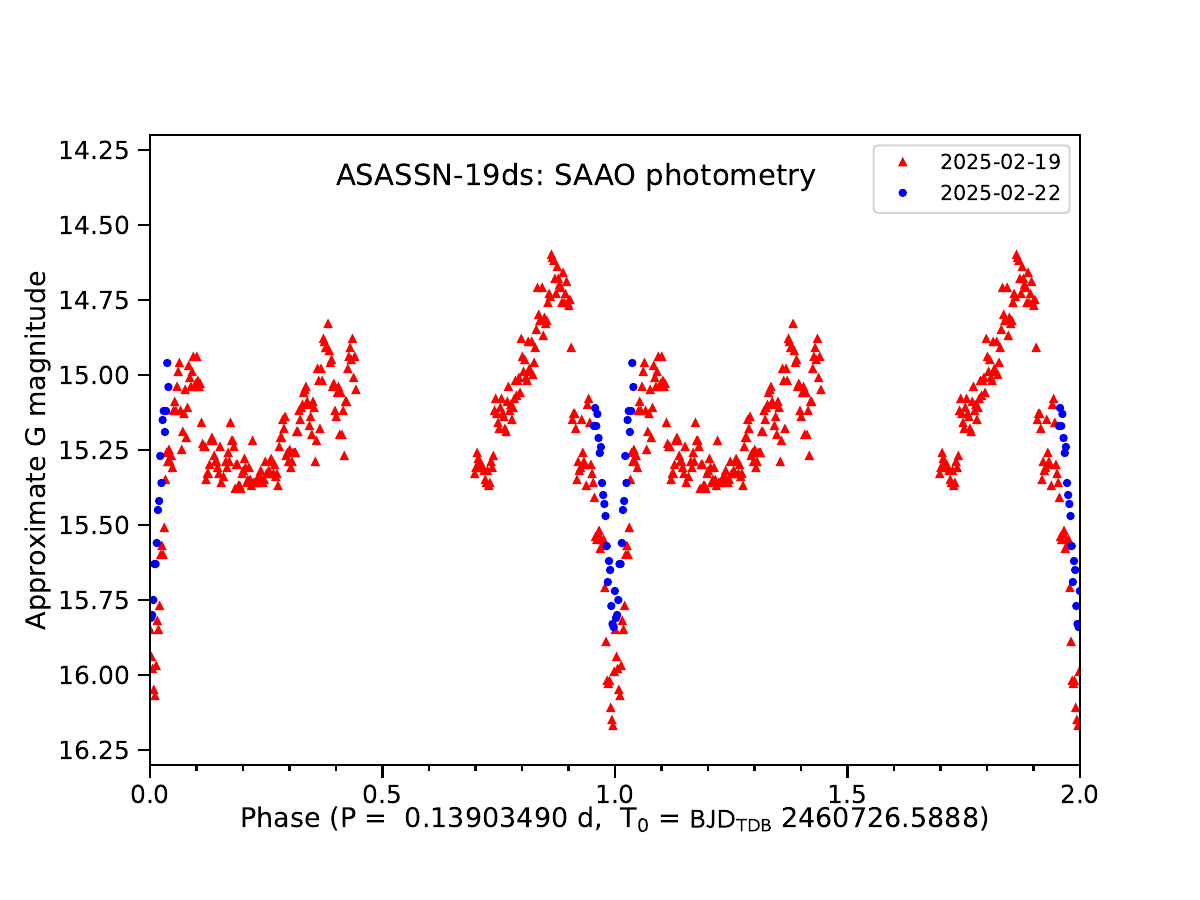}
\vskip-0.2in
\caption{
SAAO light curve of \AS\ on two nights in 2025 February, with orbital phases calculated from the ephemeris given on the $x$-axis label. The star shows an eclipse about 1~mag deep, and rapid flickering outside eclipse.
\label{fig:saaophotometry}
}
\end{figure}

Figure~\ref{fig:saaophotometry} plots the resulting light curves. The orbital phases are calculated using the ephemeris derived in the next subsection. The first night covered nearly an entire orbital cycle, displaying an eclipse with a depth of about 1~mag, lower-amplitude flickering outside eclipse, { and a flare, or possible signature of a ``hot spot'' on the accretion disk, just before the onset of the eclipse.} Only the eclipse was observed on the second night.

\bigbreak
\bigbreak
\bigbreak
\bigbreak

\subsection{Archival Photometry and Eclipse Ephemeris}

We refined the ephemeris by
estimating the midpoints of the two eclipses in our
time-series photometry, and combining these
timings with those we derived from archival survey photometry from the following three sources:
(1)~The field of the object was observed continuously in 2023 for an interval of 26.5~days by the {\it Transiting
Exoplanet Survey Satellite\/} (\TESS; \citealt{Ricker2015}),
in Sector~63. We downloaded the 120~s cadence Science Processing Operations Center
(SPOC) data using the {\tt python} package {\tt lightkurve} \citep{Lightkurve2018}.\footnote{The \TESS\/ data for ASASSN-19ds can be obtained from the Mikulski
Archive for Space Telescopes (MAST) at the Space Telescope Science Institute, at \dataset[doi:10.17909/qbfc-9h26]{https://doi.org/10.17909/qbfc-9h26}}   (2)~We obtained Asteroid Terrestrial-impact Last Alert System (ATLAS) data \citep{Tonry2018},
using their forced-photometry server \citep{Shingles2021}, and included 2,284 measurements
spanning from 2015 to 2025 in the final sample.
(3)~Finally, we downloaded data from ASAS-SN,\footnote{\url{https://asas-sn.osu.edu/}}
which
yielded 4,680 points from 2014 to 2025.

The eclipse timings derived from these large data sets yielded a unique
eclipse ephemeris:
$$
t_{\rm eclipse} = {\rm BJD_{TDB}} \, 2460726.5888(1) + 0.13903490(2)\, E\, ,
$$
where the numbers in parentheses are the estimated
uncertainties in the final digits.
Note that the Barycentric Dynamical Time (TDB) system used here is
a uniform time scale, unlike Coordinated Universal Time (UTC); TDB is
at present about 69.18~s ahead of UTC\null.  \TESS\/
data are in TDB, and we converted the other data, originally
in UTC, to TDB prior to analysis.
We refined the period and epoch for each data set
using an interactive program that plots the data
folded on a trial period, and allows the user to adjust
the period and epoch.  We assigned
cycle numbers to the data in each set---these
were completely unambiguous---and fitted a linear
ephemeris to all of the eclipse times.  The residuals from the linear fit given above were
all less than 20~s.

\begin{figure}[h]
\centering
\includegraphics[width=0.47\textwidth]{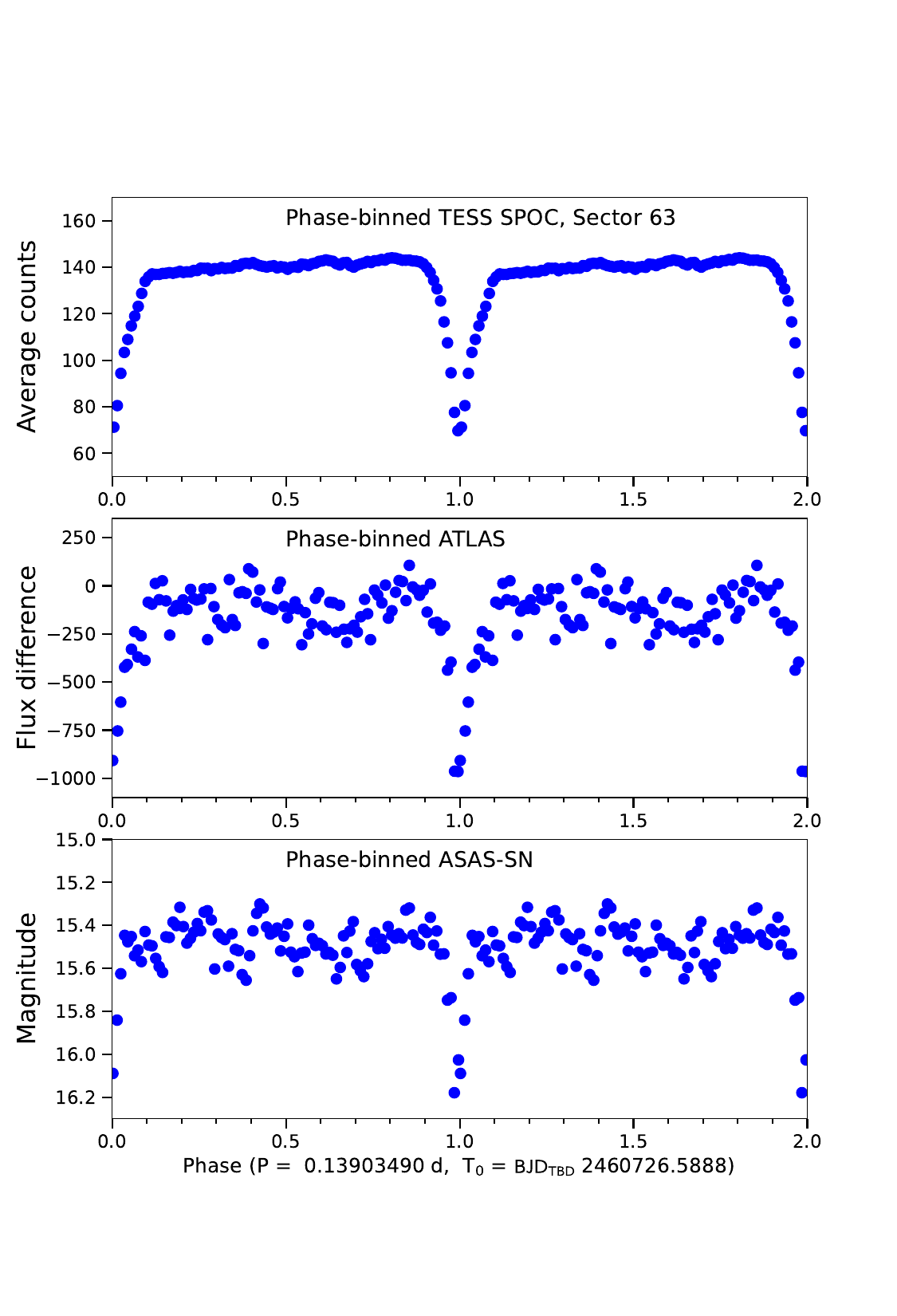}
\vskip-0.5in
\caption{
Phase-binned light curves of \AS\ from \TESS\/ (top panel), ATLAS (middle panel), and ASAS-SN (bottom panel). 
\label{fig:phasebinned}
}
\end{figure}

We averaged the data from the three archival sources into 100 phase bins per orbit, using the above ephemeris. The resulting phased light curves are shown in Figure~\ref{fig:phasebinned}. Here the rapid flickering, and the variations with timescales of weeks described below, are largely averaged out, producing a nearly flat light curve outside the sharp eclipse. This is a typical light curve for a CV with a luminous accretion disk seen approximately edge-on. It is very similar, for example, to those of the well-known eclipsing NLVs RW~Trianguli \citep[e.g.,][]{Subebekova2020} and UX Ursae Majoris \citep[see][]{deMiguel2016}.  

\bigbreak
\bigbreak
\bigbreak
\bigbreak


\subsection{ASASSN-19ds as a Z Cam Variable}

ZCVs typically show regularly spaced DN-like brightenings (except when they enter into standstills), giving their light curves a ``sawtooth'' appearance (see, for example, the collections of ZCV light curves in Figure~2 of \citealt{Shafter2005}, and in Section~3 of \citealt{Simonson2014}). The time intervals between successive maxima of a system are variable, and are generally in the range of about 10 to 30~days. 

\begin{figure}[h]
\centering
\includegraphics[width=0.47\textwidth]{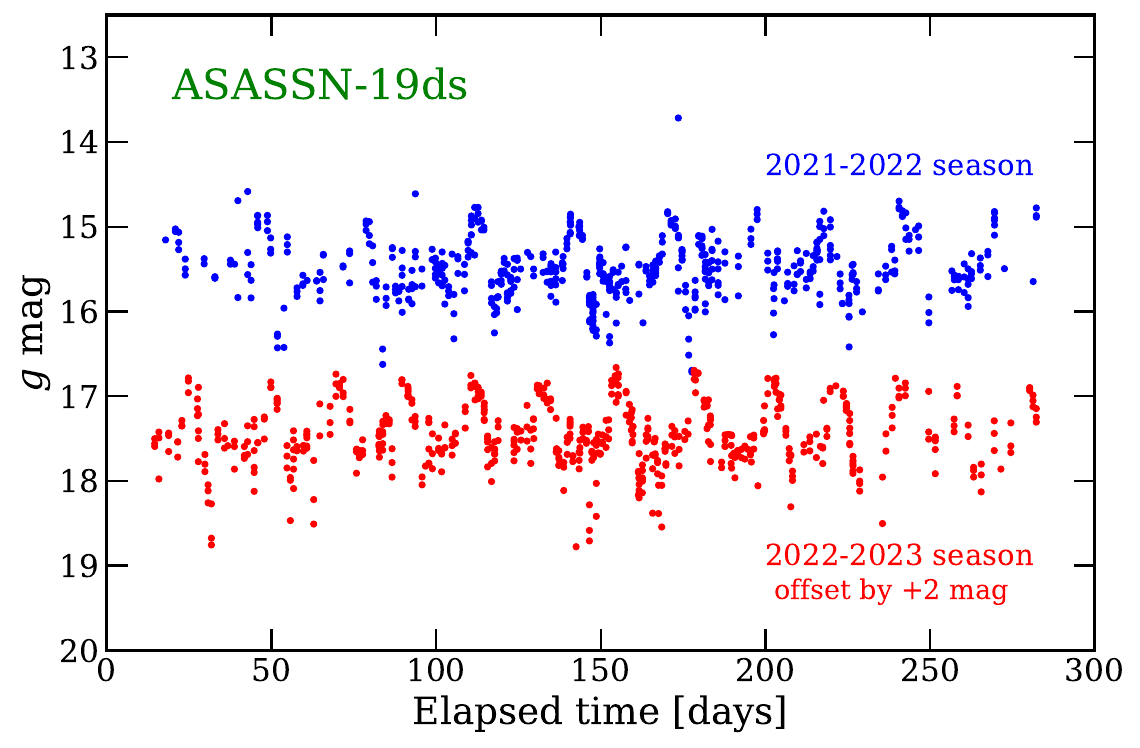}
\caption{
Seasonal $g$-band light curves of ASASSN-19ds from data in the ASAS-SN archive. Blue points are for 2021 October~30 to 2022 July~21, and red points (offset by +2~mag) are for 2022 November~16 to 2023 August~10. The star shows a ``sawtooth'' light curve, typical of Z~Cam variables, with a variable time interval between successive brightness maxima. 
\label{fig:asassn_seasons}
}
\end{figure}

ASASSN-19ds exhibits this characteristic behavior. As examples, Figure~\ref{fig:asassn_seasons} plots light curves from two seasons of photometry from the ASAS-SN monitoring. The star undergoes regularly spaced brightenings, but they are not at a precisely fixed period. At the beginning of the 2021 October to 2022 July observing season, the maxima were separated by about 34.4~days, but this interval shortened, and by the end of the season the spacing was about 24.8~days. For the second season, 2022 November to 2023 August, the interval between consecutive maxima was nearly constant, at about 21.8~days.

As discussed in our Paper~I, the ZCV SY~Cnc exhibited very similar regular outbursts, in fact at nearly the same range of intervals between eruptions as seen in ASASSN-19ds. We also note that the nominal absolute magnitude of ASASSN-19ds, $M_G=+6.5$ (Table~\ref{tab:DR3data}), is quite similar to those derived from \Gaia\/ DR3 data for the ZCVs Z~Cam (+6.0) and AT~Cnc (+6.2). 

In summary, all of the evidence points to a classification of ASASSN-19ds as belonging to the ZCV subcategory. However, it has one of the shortest known orbital periods among ZCVs---shorter, for example, than for all but one (BX~Puppis) of the 22 ZCVs listed by \citet{Shafter2005} and \citet{Simonson2014}, which range from 0.127 to 0.380~days.


\section{Deep Imaging \label{sec:deep_imaging} }

Deep imagery of the nebula surrounding ASASSN-19ds was obtained by Stern, using his PlaneWave CDK24 0.6\,m $f/6.5$ reflecting telescope\footnote{\url{https://planewave.com/products/cdk24-ota}} on a PlaneWave L-600 mount,\footnote{\url{ https://planewave.com/products/l-600-telescope-mount}} equipped with a Moravian Instruments C5-100 CMOS camera.\footnote{\url{https://www.gxccd.com}} The telescope is located at the dark-sky site of Deep Sky Chile,\footnote{\url{https://www.deepskychile.com}} in R\'io Hurtado Municipality, Chile, about 39~km south of Cerro Tololo Interamerican Observatory.
Observations were carried out between 2025 January~28 and February~6. Exposures were 300~s each, with total numbers of exposures of 35, 30, 34, 235, and 331, respectively, in R, G, B, \Ha, and [\oiii] filters manufactured by Chroma.\footnote{\url{https://www.chroma.com/}} The total integration time was 55.4~hours. 

\begin{figure*}[h]
\centering
\includegraphics[width=6in]{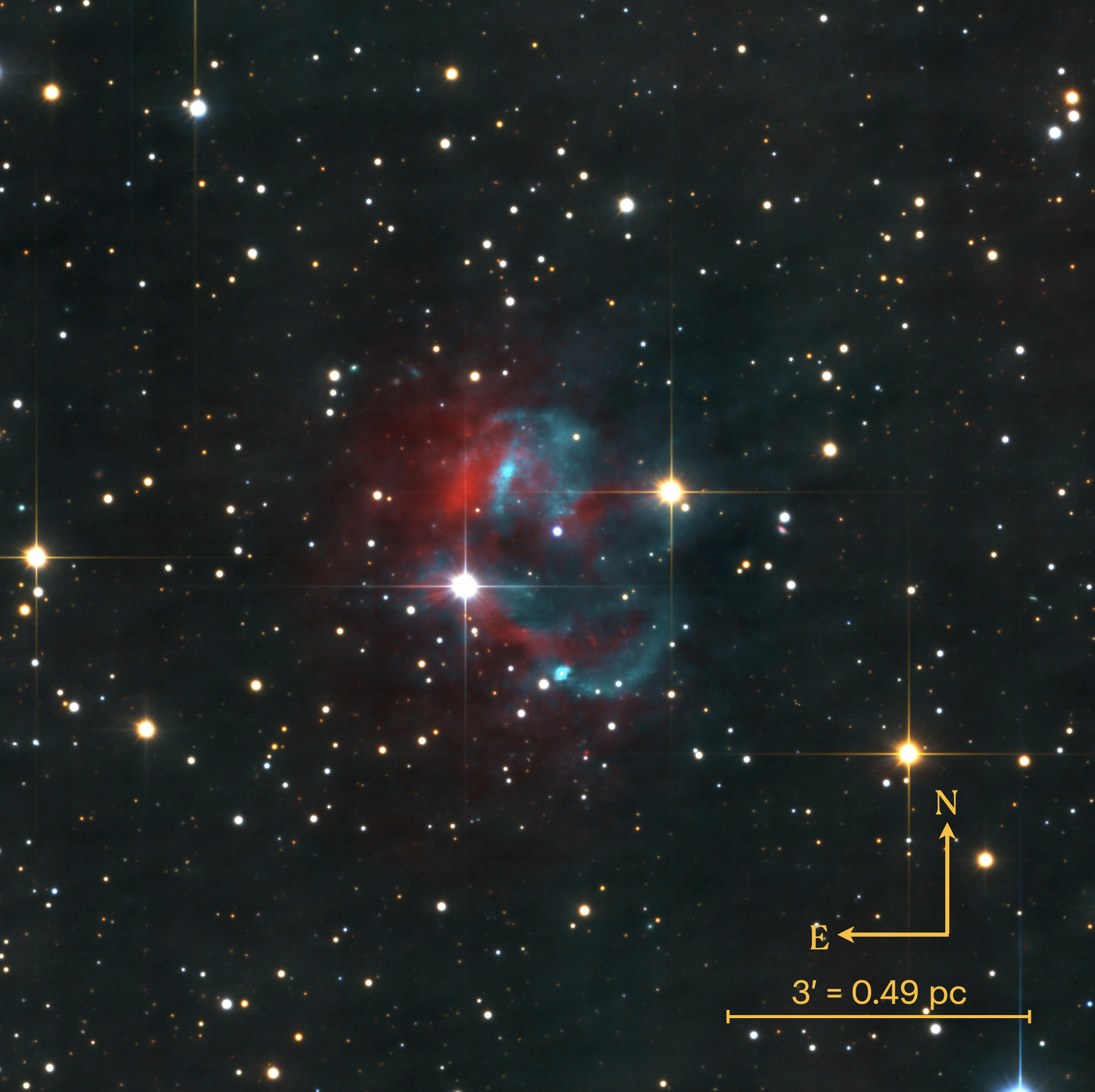}
\caption{
Image of the nebula around \AS, created from 55.4~hours of exposure time in RGB, \Ha, and [\oiii] $\lambda$5007 filters as described in the text. \Ha\ is assigned to the red channel, and [\OIII]  to the green and blue channels. Orientation and scale of the image (including a conversion to linear size at the distance of the star) are indicated at the lower right.
\label{fig:deepimage}
}
\end{figure*}

Figure~\ref{fig:deepimage} displays a color image of the nebula,
created by Stern from these frames using the
PixInsight tool. An ``HOO'' palette was employed, with \Ha\ assigned to the red channel, and [\OIII] $\lambda$5007 to the green and blue channels, while the RGB bandpasses were used for the surrounding stellar field.
 
In the light of [\oiii], the nebula has a mostly hollow and tattered elliptical morphology, with minor and major axes of roughly $0\farcm59 \times 0\farcm96$, or about $0.29 \times 0.47$~pc at the distance of ASASSN-19ds. The \Ha\ emission is more diffuse, and is brightest on the northeast side of the star.



\section{Nature of the Nebula \label{sec:nature_of_nebula} }

We consider three possible interpretations of the nebulosity around ASASSN-19ds: (1)~result of an encounter of a UV-bright CV with an interstellar cloud; (2)~ejecta from a nova outburst; or (3)~planetary nebula.

\smallbreak

1. {\it Chance encounter.} As discussed in our introduction, in Papers~I and~II we presented images of faint nebulae discovered around three CVs: SY~Cnc (Paper~I) and LS~Peg and ASASSN-V J205457.73+515731.9 (Paper~II)\null. All three exhibit bow shocks, prominent in [\oiii] $\lambda$5007, on the leading edges of the stars in the directions of their proper motions. Moreover these three new examples join two further CVs with similar bow shocks: BZ~Cam and V841~Ara (see discussion and references in Papers~I and~II). 

Creation of a bow shock requires a large relative motion between a star, which must emit a fast wind, and the surrounding relatively stationary ISM\null. We argued for a scenario in which these five NLVs and ZCVs (all of which, as shown by their absolute magnitudes, have luminous accretion disks, known to launch fast winds into space) are by chance passing through interstellar clouds that are sufficiently dense for production of bow shocks. 

Deep imagery also shows that these stars are located near the edges of large, faint \Ha-emitting nebulae. We interpreted these off-center nebulae as being material photoionized and/or collisionally excited by the passage of UV- and X-ray-bright CVs and their stellar winds through the ISM\null. As the stars move on, they leave behind ``recombination wakes,'' in the form of the off-center nebulae seen in \Ha\ filters (which also have transmission at the neighboring low-ionization [\nii] lines at $\lambda\lambda$6548 and 6583). Another case of this phenomenon was discussed by \citet{Bond2023Fr2-30}, who presented deep imaging by Talbot of the nebula Fr~2-30; it surrounds a hot subdwarf~O star that is unlikely to have recently ejected material. Instead, it lies in a field covered with faint nebulosity, and clearly appears to be leaving behind a trail of recombining photoionized material.

However, we have doubts that ASASSN-19ds is a similar case of a random encounter. Its proper motion, based on the \Gaia\/ DR3 data in Table~\ref{tab:DR3data}, and corrected\footnote{This adjusts the proper motion to make it relative to the standard of rest at the star's location. We used a {\tt python} code created by S.~del Palacio (see \citealt{Martinez2023}), available at \url{https://github.com/santimda/intrinsic_proper_motion}} for Galactic rotation, is $2.61\pm0.10\,\masyr$ at position angle $264^\circ$. This motion is almost directly to the right in Figure~\ref{fig:deepimage} and in the { pair of figures} presented below. We see no evidence for a bow shock in that direction.  Moreover, the nebula has a morphology consistent with having been ejected from the star, as we now discuss.


\smallbreak

\begin{figure}[h]
\centering
\includegraphics[width=0.47\textwidth]{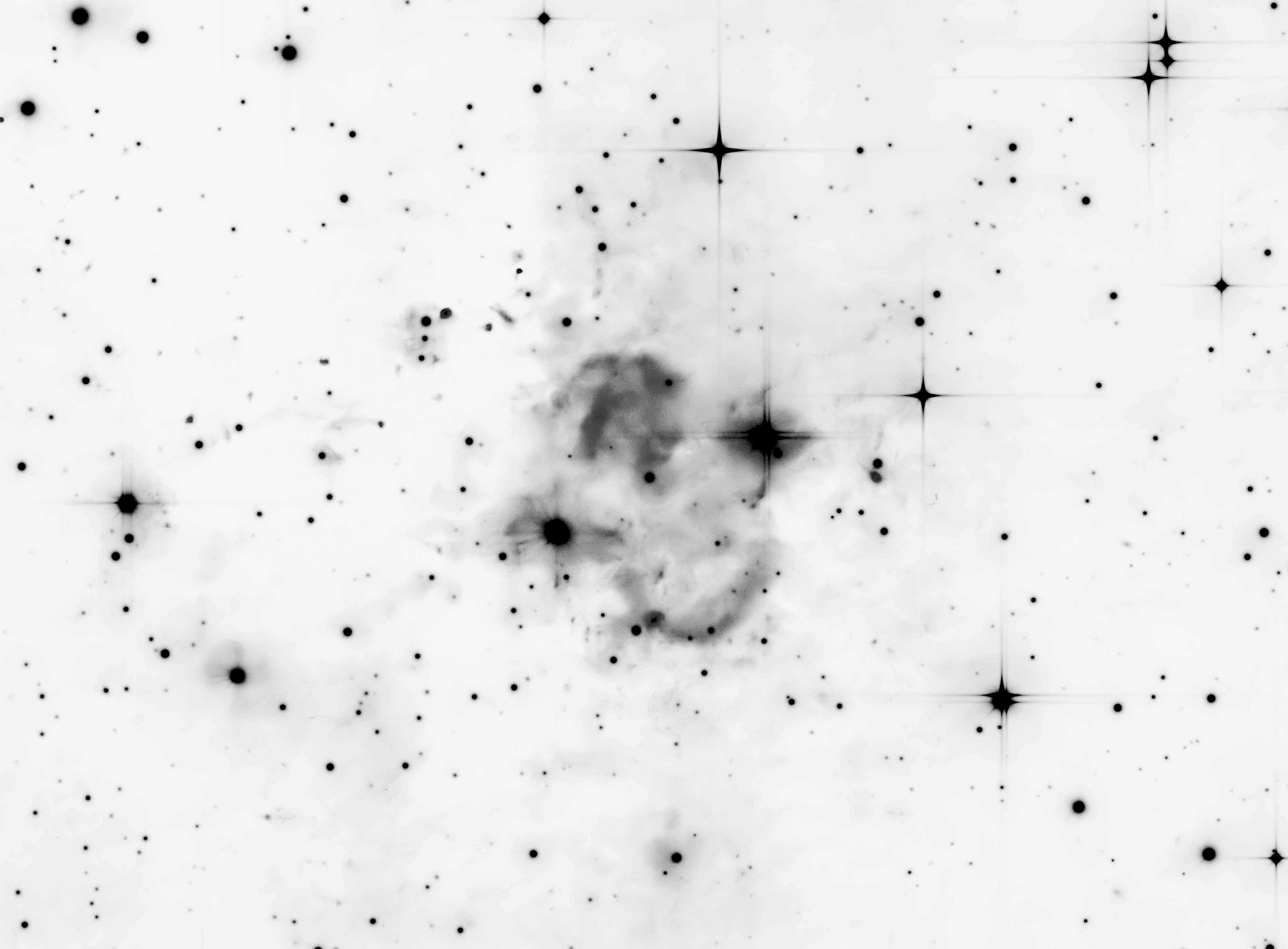}
\vskip0.02in
\includegraphics[width=0.47\textwidth]{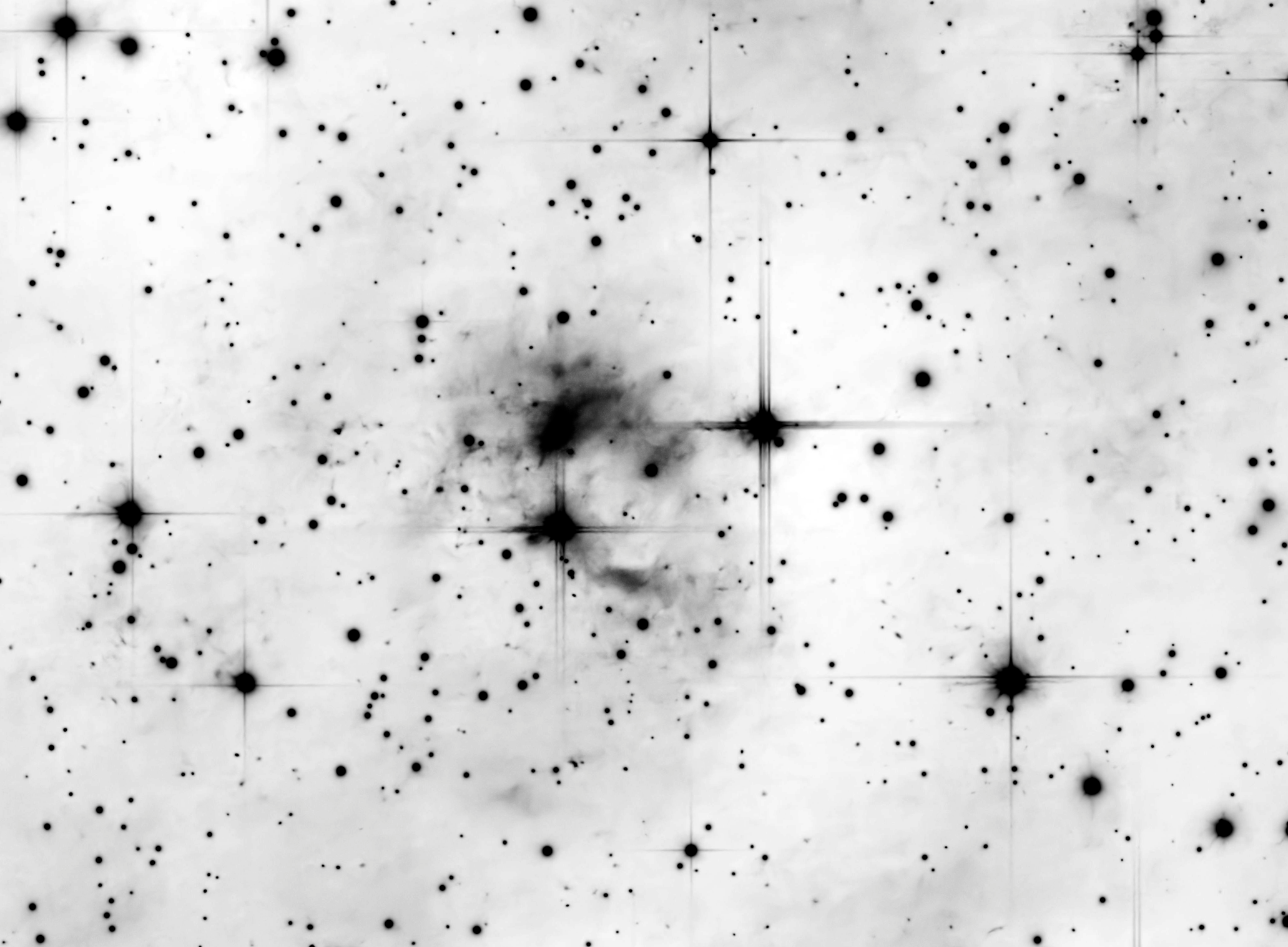}
\caption{
{ Deep images of the \AS\ nebula in the light of [\oiii] $\lambda$5007 (top frame) and of \Ha\ (bottom frame). }
\label{fig:stern_oiii}
}
\end{figure}

2. {\it Nova remnant.} Is the ASASSN-19ds nebula material ejected from an ancient (and unrecorded) nova eruption? { We prepared  images of the nebula from Stern's data in [\oiii] and \Ha, stretched to show the faintest features more clearly than in the multicolor picture above. They are shown in Figure~\ref{fig:stern_oiii}. The morphology seen here in the [\oiii] frame is suggestive of a bipolar outflow along an axis passing through the star from the north-northeast to the south-southwest. At the two ends of the axis there are nebulous arcs, likely indicating a deceleration as the outflow encounters the ISM, as further discussed below. The presence of these features in [\oiii], but not in the \Ha\ frame, is consistent with collisional excitation.}  

In a recent paper, \citet[][hereafter S25]{Santamaria2025} have presented an atlas of images of 66  remnants of known nova eruptions. Large fractions of the sample of resolved nebulae are circular (27\%) or elliptical (56\%), and in most cases the rings and ellipses are thin and relatively intact, surrounding hollow interiors. The ASASSN-19ds nebula does not have a structure of this type. A handful of remnants show a myriad of small knots, but this feature is not seen in ASASSN-19ds either. Only a very few of the nova shells in the S25 atlas have a bipolar morphology somewhat similar to that of ASASSN-19ds.  

S25 show that the physical radii, $r$, of nova remnants are statistically proportional to age, $\tau$, with a mean relation of $(r/{\rm pc}) = (0.000725 \pm 0.000125) \, (\tau/{\rm yr})$. Assuming that the ASASSN-19ds nebula is a nova remnant, with $r\simeq0.19\,\rm pc$ (mean of the semi-minor and semi-major axes from Section~\ref{sec:deep_imaging}), this relation implies an age of roughly 260~yr with an uncertainty of at least $\pm$40~yr---with the caveat that this estimate is based on an extrapolation of the S25 relation to ages considerably greater than those of the calibrators. 

The visual absolute magnitudes of CNe at maximum light average about $-7.5\pm1.0$~mag \citep[e.g.,][and references therein]{Selvelli2019}, with a few outliers brighter than about $-9$. Thus, if ASASSN-19ds had undergone a nova explosion, its apparent magnitude at maximum would have been about $1.5\pm1.0$~mag, or conceivably slightly brighter than magnitude zero. This makes it somewhat puzzling that no such spectacular event,\footnote{ASASSN-19ds is nearer than all but six of the 66 old novae listed by \citet{Santamaria2025}} occurring around the middle of the 18th Century, has (to our knowledge) been recorded. However, it is certainly possible that, had it occurred, it was missed because of its southerly declination, and\slash or the Sun being too close.

The faint nebulosity around Z~Cam itself, mentioned in Section~\ref{subsec:faint_nebulae}, is not included in the S25 atlas, but deep narrow-band images have been presented by \citet{SharaZCam2012,  Shara2024}. The Z~Cam nebulosity has several features in common with that around ASASSN-19ds. In particular, \AS\ in [\oiii] shows a prominent thin arc on its south-southwest side, and the Z~Cam nebula exhibits a remarkably similar structure, also to its south-southwest. (Another similar set of arcs is seen adjacent to the symbiotic nova RX~Puppis in images published by \citealt{Ilkiewicz2024}; and somewhat similar features are seen in the nebula IPHASX J210204.7+471015, whose central star is a nova-like CV, discussed by \citealt{Guerrero2018}.) Short segments of arcs are also visible in [\oiii] on the north-northeast side of \AS. In the references cited, such arcuate features are attributed to fast ejecta from the central objects decelerating as they ``snowplow'' into the surrounding ISM\null. The same interpretation is plausible for \AS, and is consistent with it having undergone a CN outburst that launched the ejecta a few centuries ago.



Deceleration of expanding nova shells due to the ISM was discussed by \citet{Oort1946}, who predicted a deceleration timescale that could be as short as about a century. \citet{Duerbeck1987} measured decelerations with timescales of $\sim$75~yr in four nova remnants; however \citet{Santamaria2020} studied shells around five old novae (including one also measured by Duerbeck) and concluded that all five are still expanding freely after intervals as large as 130~yr after the outbursts. It seems clear that the time to the onset of deceleration varies among different objects, depending on parameters of the explosions and the density of the surrounding ISM\null. But in any case, \AS\ appears to be well into the deceleration phase, as indicated by its tattered appearance and the arcs at the ends of the major axis---if indeed the nebula is material ejected during a nova outburst.

The age of the ASASSN-19ds nebula estimated above makes it older than almost all of the nova remnants in the S25 atlas. These objects have a median age of only 33~yr, and very few are older than $\sim$100~yr. If the deceleration of \AS\ is well underway, the nebula's age could be significantly higher than given above by the linear relation.

\smallbreak

3. {\it Planetary nebula.} Finally we consider the possibility that the \AS\ nebula could be an ancient planetary nebula (PN), rather than the remnant of a recent nova explosion. As \citet{Miszalski2016} remark,\footnote{This remark is made in a paper by the cited authors on the faint nebula Te~11, whose central star is an eclipsing DN with an orbital period of 0.121~days. They consider the possibility that the nebula is a PN, but conclude that it is more likely an ancient nova remnant. The Te~11 nebula has a bipolar morphology somewhat similar to that of \AS.} ``The appearance of a nova inside a PN would be an oddity, but is not unprecedented.''

Among the nova remnants illustrated in the S25 atlas, the nebulosity around V458~Vulpeculae appears most similar to that of \AS. The V458~Vul nebula shows an elongated and hollow elliptical structure, with gaps at both ends of the minor axis---a morphology suggestive of an earlier evolutionary stage of the more evolved and expanded \AS\ nebula. Shortly after the eruption of V458~Vul in 2007, \citet{Wesson2008} showed that its nebulosity had been in place before the outburst. Moreover, these authors found a full width at half maximum for nebular emission lines of only $13.7\,\kms$, much lower than the line widths found in expanding nova remnants. Thus the nebula is not material ejected during either the 2007 eruption, nor in a previous nova explosion. Instead, it is a pre-existing planetary nebula (PN)---which was seen to brighten considerably as it was ``flash-ionized'' by UV emission from the 2007 nova explosion, as described by \citet{Wesson2008}. 

In a follow-up spectroscopic investigation, \citet{Rodriguez2010} established that V458~Vul is a CV with an orbital period of only 0.068~days (1.63~hr). They presented a scenario in which the system has undergone two common-envelope episodes, and suggested that the total mass of the binary is more than the Chandrasekhar mass, making it a candidate Type~Ia supernova progenitor.


Could \AS\ be a similar case of a short-period CV lying inside an ancient PN, ejected as a result of common-envelope interaction(s)? In the absence at present of information  on the expansion velocity of the nebula, this possibility remains a viable---and testable---alternative to it being ejecta from a relatively recent nova eruption.

\section{Summary and Future Work}

To summarize our findings, we have discovered a faint bipolar nebulosity surrounding the previously little-studied CV \AS. Based on archival and newly obtained photometry, we confirm that the star is an eclipsing binary with an orbital period of 0.139~days (3.34~hr). Spectroscopic observations, along with the object's absolute magnitude, and the presence of a ``sawtooth'' light curve (with a spacing that varies from about three to nearly five weeks), place the star in the Z~Cam subclass of CVs.  

We consider three scenarios for the origin of the nebulosity around \AS. In our two previous papers (Papers~I and~II) on newly discovered faint nebulae around three CVs, we proposed that they are the results of chance encounters of UV-emitting stars with interstellar clouds. However, \AS\ lacks the bow shocks in the direction of their motion seen in these objects; instead, the bipolar morphology is strongly suggestive of ejection of nebular material from the star itself.

The \AS\ nebula may be the remnant of an unobserved nova outburst that occurred some two and a half (or more) centuries ago; or it could conceivably be a much older PN\null. We note that bipolar morphologies such as seen around \AS\ are rare among nova remnants, but common in PNe. To decide between these alternatives, we urge spectroscopy of the nebula, since the expansion velocities in nova remnants are much higher than in PNe and should be decisive. Moreover, it would be useful to measure proper motions in the nebulosity, which could also help make this distinction. However, such investigations would be challenging because the nebulosity is extremely faint.

\bigbreak
\bigbreak

\acknowledgments



H.E.B. remembers with fondness his teacher Dean B. McLaughlin (1901--1965), a leading authority on classical novae, who would have been so amazed and delighted to see images like those presented here.

The Digitized Sky Surveys were produced at the Space Telescope Science Institute under U.S. Government grant NAG W-2166. The images of these surveys are based on photographic data obtained using the Oschin Schmidt Telescope on Palomar Mountain and the UK Schmidt Telescope. The plates were processed into the present compressed digital form with the permission of these institutions. 

The Southern H-Alpha Sky Survey Atlas (SHASSA) was supported by the National Science Foundation.

This work has made use of data from the European Space Agency (ESA) mission
{\it Gaia\/} (\url{https://www.cosmos.esa.int/gaia}), processed by the {\it Gaia\/}
Data Processing and Analysis Consortium (DPAC,
\url{https://www.cosmos.esa.int/web/gaia/dpac/consortium}). Funding for the DPAC
has been provided by national institutions, in particular the institutions
participating in the {\it Gaia\/} Multilateral Agreement.



Based in part on observations made with the NASA {\it Galaxy Evolution Explorer.}
\GALEX\/ was operated for NASA by the California Institute of Technology under NASA
contract NAS 5-98034.  

This research has made use of the SIMBAD and Vizier databases, operated at CDS, Strasbourg, France.

This paper uses observations made at the South African
Astronomical Observatory (SAAO), taken as part of the 2025 Dartmouth
Foreign Study Program in Astronomy.  We thank Professor Ryan
Hickox and graduate student Emmanuel Durodola for supervising
the time-series photometric measurements.  Assistance was provided by Dartmouth undergraduates
Aryan Bawa, Michael Farnell, Shreya Gandhi, Piper Gilbert,
Gavin Goss, Kushal Jayakumar, Alexandra Lipschutz, Timothy
McGrath, Ricardo Mendez, Annabelle Niblett, Kate Schwendemann,
Beatrice Sears, Arnav Singh, Aimilia Tsopela, Divik Verma,
Kendall Yoon, and Lauren Zanarini.

This paper includes data collected by the \TESS\/ mission.
Funding for the \TESS\/ mission is provided by the NASA Science
Mission Directorate.  

This work has made use of data from the Asteroid Terrestrial-impact Last Alert
System (ATLAS) project. The ATLAS project is primarily funded to search for near-Earth asteroids through
NASA grants NN12AR55G, 80NSSC18K0284, and 80NSSC18K1575; byproducts of the NEO
search include images and catalogs from the survey area. This work was
partially funded by Kepler/K2 grant J1944/80NSSC19K0112 and HST GO-15889, and
STFC grants ST/T000198/1 and ST/S006109/1. The ATLAS science products have been
made possible through the contributions of the University of Hawaii Institute
for Astronomy, the Queen's University Belfast, the Space Telescope Science
Institute, the South African Astronomical Observatory, and The Millennium
Institute of Astrophysics (MAS), Chile.


We thank S.~del Palacio, author of the {\tt python} code used in Section~\ref{sec:nature_of_nebula}, for pointing out to us the importance of correcting proper motions for the effect of differential Galactic rotation, and for kindly assisting us in implementing his code.




\bibliography{PNNisurvey_refs}

\end{document}